# Enhancement of Exciton Valley Polarization in Monolayer MoS$_2$ Induced by Scattering

*Yueh-Chun Wu,[1] Takashi Taniguchi,[2] Kenji Watanabe[3] and Jun Yan[1,*]*

[1]Department of Physics, University of Massachusetts Amherst, Amherst, Massachusetts 01003, USA

[2] International Center for Materials Nanoarchitectonics, National Institute for Materials Science, Tsukuba, Ibaraki 305-0044, Japan

[3] Research Center for Functional Materials, National Institute for Materials Science, Tsukuba, Ibaraki 305-0044, Japan

[*]Corresponding Author Email: yan@physics.umass.edu

**Abstract:** We report on scattering induced valley polarization enhancement in monolayer molybdenum disulfide. With thermally activated and charge doping introduced scattering, our sample exhibits seven- and twelve-folds of improvements respectively. This counter-intuitive effect is attributed to disruptions to valley pseudospin precession caused by rapid modulation of exciton momentum and concomitant local exchange interaction field, at time scales much shorter than the precession period. In contrast, the valley coherence is improved by thermally activated scattering, but not by charge doping induced scattering. We propose that this is due to anisotropic pseudospin scattering and generalize the Maialle-Silva-Sham model to quantitatively explain our experimental results. Our work illustrates that cleaner samples with minimal scattering, such as those carefully suspended or protected by hexagonal boron nitride, do not necessarily lead to good valley polarization. Well-controlled scattering can in fact provide an interesting approach for improving valleytronic devices.

The advent of two-dimensional materials, such as graphene and transition metal dichalcogenides (TMDs), has boosted the development of valleytronics, devices that harness the valley degree of freedom [1–3]. In monolayer (1L) TMD semiconductors, there exists two energetically degenerate valleys +K and -K, where the conduction and valence bands are separated by about 2 eV. Governed by the symmetry of the two valleys, circularly polarized light near the optical bandgap only promotes electrons in one valley from the valence band to the conduction band, and is forbidden to couple to the other valley [3]. This provides a mechanism to generate valley polarization by optical excitation (Fig.1a). Subsequent radiative recombination of bound electron-hole pairs, i.e. the excitonic photoluminescence (PL) emission, provides information on the degree of valley polarization in the atomic layer, via radiation intensity difference in $\sigma^+$ and $\sigma^-$ polarization channels, corresponding to bright excitons originated from the +K and -K valleys [4–6].

An important task in TMD valleytronics is to identify mechanisms that drive valley depolarization and approaches that can be used to improve valley polarization. However, despite several years of intensive investigation, how optically populated excitons in 1L-TMDs lose their valley polarization is still controversial. PL excitation (PLE) studies that reveal roll off of valley polarization have been interpreted with two different mechanisms: one is the intervalley emission of two K-point LA phonons by excitons with high kinetic energy [7,8], and the other is the valley-depolarizing exchange interaction [9–11] that becomes stronger for excitons with higher center-of-mass momentum [12]. Intervalley exchange interaction was also employed to explain the temperature dependence of exciton valley polarization [13]; but other studies attributed this to thermally activated intervalley scattering by modes such as large momentum phonons [5,14]. More recent theoretical studies further propose that dark excitons can play a role in bright exciton valley depolarization [15,16].

In this letter, we report a counter-intuitive scattering-induced enhancement of valley polarization. This observation provides pivotal evidence for pseudospin precession under effective fields (Fig.1b) in 1L-TMDs. The valley pseudospin precession becomes significantly disrupted in the presence of frequent exciton momentum scattering, at a rate that is faster than the precession frequency. We further discovered that pseudospin scattering is anisotropic under charge doping, an effect that can be traced to valley-selective exciton-polaron dressing. A generalized Maialle-Silva-Sham (MSS) model [10] is developed to explain our experimental results quantitatively. These studies offer key insights into many mysteries behind valley depolarization in 1L-TMDs.

Our device structure is schematically shown in Fig 1c. The 1L-$MoS_2$ sample is exfoliated from bulk crystals and sandwiched between two hexagonal boron nitride (hBN) flakes using a dry transfer technique [17–19]. The atomic stack is then used to pick up a few-layer graphene flake, which serves as the back gate. The device is mounted inside a cryostat with optical access and cooled down to a base temperature of 3.4 K. A 633 nm (1.96 eV) He-Ne laser is used to generate excitons in the sample.

Figure 1d top panel shows circular polarization resolved PL of our device at zero gate. The hBN sandwiched 1L-MoS$_2$ is of decent quality, with the neutral exciton PL at 1.943 eV exhibiting a full width at half maximum (FWHM) of 2.5 meV, much narrower than typical devices exposed to air and is comparable to MoS$_2$ samples of high quality reported so far [20–22]. The valley polarization $P = \frac{I_{\sigma+} - I_{\sigma-}}{I_{\sigma+} + I_{\sigma-}}$ is quite small, only 0.05 despite good sample quality. It is indeed worth noting that higher quality sample with narrower PL linewidth does not necessarily translate to good valley polarization. Samples with relatively large inhomogeneous broadening can reach quite high *P* [6,12,23,24], compared with 'better' samples protected by hBN [20–22,25]. We briefly comment that this observation is in line with the main spirit of our work here: well-controlled scattering can provide an effective mechanism to improve valley polarization in 1L-TMDs [24]. In Fig.1d, we additionally measured linear polarization resolved PL and found valley coherence [26] $C = \frac{I_{\sigma x} - I_{\sigma y}}{I_{\sigma x} + I_{\sigma y}} = 0.17$, also a relatively small value consistent with the small *P*.

The key experimental observation of our paper is presented in Fig.2: raising the sample temperature to about 50K (Fig.2a) and introducing charge doping (Fig.2c) can markedly enhance *P*, from 0.05 to 0.35 (Fig.2b) and to 0.6 (Fig.2d) respectively. We first discuss the temperature dependence. From 3.4K to 200K, *P* increases as T goes to 50K and then drops to near 0 at higher temperatures. The non-monotonic variation is in contrast to the monotonic broadening of linewidth (Fig.2b) and redshift of peak energy (Supplementary [27] Fig.S3c). This immediately rules out either K-point phonon scattering or exchange interaction strength variance as the single dominant valley depolarization mechanism, since they both predict monotonic decrease of valley polarization [7,12,13]. In fact, at 3.4K our laser is detuned only 17meV away from the bright exciton energy, much smaller than the energy of two K-point phonons [28,29]. Note that the valley polarization enhancement is not due to resonance effects given that our PLE measurements show no resonant structure between 15 - 32 meV detuning, and that measurements with constant laser detuning give almost identical temperature dependence for the valley polarization; see Supplementary [27] S7 and Fig.S5.

Most remarkable in our data is the behavior between 3.4 and 50 K. Naively thermally activated scattering that broadens the PL linewidth is anticipated to undermine device performance such as valley polarization, but experimental results tell the opposite. This scattering enhanced valley polarization is reminiscent of motional narrowing in nuclear magnetic resonance (NMR) spectroscopy [30], or spin relaxation via the D'yankonov-Perel' (DP) mechanism [31], where precession of nuclei or electron spin under real or effective fields gets disrupted by frequent scattering at rates much higher than the precession frequency, elongating spin lifetimes.

In our case, the effective field under which TMD valley pseudospins precess is provided by the exchange interaction. Consider an exciton with wavevector $\vec{k}$ making an angle $\theta$ with the $k_x$-axis, and valley pseudospin angular momentum $\frac{\hbar}{2}\vec{\sigma}$, where $\vec{\sigma} = [\sigma_x, \sigma_y, \sigma_z]$ denotes Pauli matrices acting on the valley pseudo-spin 2D Hilbert space. The long-range exchange interaction is of the form [9–11,23,25],

$$H_{ex} = -J(k)[\cos(2\theta)\sigma_x + \sin(2\theta)\sigma_y] \quad (1).$$

Its strength $J(k) = j\frac{k}{K}$, where $j \approx 0.5 - 1\,\text{eV}$ (Supplementary [27] S3) and $K = \frac{4\pi}{3a} = 1.33\,\text{Å}^{-1}$, is proportional to exciton momentum $k = |\vec{k}|$. The presence of only $\sigma_x$ and $\sigma_y$ in the expression implies that the interaction mimics the impact of an in-plane magnetic field. Rewriting Eqn.(1) as $H_{ex} = -\vec{\mu} \cdot \vec{F}(\vec{k})$, we can interpret $\vec{\mu} = \mu\vec{\sigma}$ as a pseudo magnetic dipole moment, and $\vec{F}(\vec{k}) = \frac{J(k)}{\mu}[\cos(2\theta), \sin(2\theta), 0]$ as a pseudo magnetic field about which the exciton valley pseudospin precesses.

In a 1L-MoS$_2$ upon $\sigma^+$ optical excitation, the valley pseudospin of photo-excited excitons is initially aligned along the +z direction. To describe subsequent exciton relaxation till eventual radiative recombination, we model the excitons as possessing an average kinetic energy $E_K$, population lifetime $\tau$ and scattering time $\tau_*$. The excitons are thus represented as occupying a ring in the momentum space (red circle in Fig.1b), and their pseudospins precess with a characteristic angular velocity $\Omega$ equal to $\frac{2J(k)}{\hbar}$. Note that although $\Omega$ is the same for all excitons in this simplified description, the precession axes depend sensitively on the direction the excitons propagate: for an exciton with wavevector $\vec{k}$ making an angle $\theta$ with the $k_x$-axis, the precession axis makes an angle of $2\theta$ (Fig.1b). As a result, the ensemble of excitons with kinetic energy $E_K$ and pseudospin along the +z direction loses valley polarization over a time scale of $\Omega^{-1}$ in the absence of scattering (assuming $\tau > \frac{1}{\Omega}$).

As we raise the temperature, thermally activated momentum scattering disrupts valley pseudospin precession and this becomes important when the scattering time $\tau_*$ is much shorter than $\Omega^{-1}$ [13,16,17]. For an exciton that experiences many scattering events before completing a 2π rotation, the cumulative precession angle $\Delta\varphi$ can be mapped to the travel distance of a one-dimensional random walk, and $\Delta\varphi^2(t) = (\Omega\tau_*)^2 \frac{t}{\tau_*}$ where $\Omega\tau_*$ is angle processed between two scattering events, and $\frac{t}{\tau_*}$ is the number of scattering events over $t$. The valley polarization is lost at the time $\Delta\varphi$ approaches unity, giving a valley depolarization time of $1/(\Omega^2\tau_*)$. In this strong scattering regime, the more frequent the scattering, i.e., the shorter the $\tau_*$, the longer it takes lose the valley polarization, similar to motional narrowing in NMR and spin relaxation via the DP mechanism [30,31]. The scattering elongated valley polarization lifetime is the root cause of our observed enhancement of valley polarization.

Before moving forward with quantitative modeling, we examine the gate dependent data in Fig.2c. Here the increase of valley polarization concomitant with broadening of linewidth (Fig.2d) is also consistent with scattering induced enhancement. Note that there is a slight blueshift of exciton energy with doping; however our PLE studies (Supplementary [27] S7; Fig.S5) confirm that this blueshift has negligible impact on valley polarization. Another plausible mechanism for *P* improvement is the reduction of exchange interaction strength due to free charge screening, an effect that was invoked to explain a recent gate dependence study of 1L-WSe$_2$ [32]. To distinguish between these two mechanisms, we performed valley coherence measurements using linearly polarized excitation and collection (PL spectra in Fig.S8; similar measurements also performed for temperature dependence). Figure 3 a&b show the evolution of valley coherence *C* for T and $V_g$ dependence. The valley polarization in Fig.2 b&d is also included for comparison. Interestingly while as a function of temperature *C* shows trends similar to *P*, the $V_g$ dependence of the two are quite different: with charge doping, *C* did not improve, and in fact, decreased slightly at high $V_g$. This observation leads us to conclude that screening is unlikely the dominant mechanism that caused *P* improvement in our device, as a decrease of exchange interaction strength should lead to similar improvements in *C*, as demonstrated in 2s exciton state studies [25].

The dichotomy between valley polarization and coherence under charge doping can be understood as due to anisotropic pseudospin scattering. The dressing of a bound electron-hole pair with a Fermi sea of charges forms exciton polarons [33–37]. In MoS$_2$ with electron-hole mass ratio close to 1, the polaron dressing occurs in an intervalley fashion: excitons in one valley is only dressed by charges residing in the opposite valley [38,39]. One can intuitively see that this type of scattering is quite detrimental to valley coherence. Consider an exciton with a wavefunction that is a coherent superposition of exciton states at +K and -K valleys, such as those generated by linearly-polarized optical fields in Fig.S8. When the exciton is scattered by a charge in +K valley, only its -K component is affected; thus, the relative phase between the +K and -K components of the wavefunction is compromised, and the valley coherence is lost.

To quantitatively explain our experimental results, we generalize the Maialle-Silva-Sham (MSS) model [10] to allow for anisotropic pseudospin scattering (see Supplementary [27] S1). The valley polarization and coherence are given by:

$$P = \frac{1}{1+\frac{4J(k)^2}{\frac{\hbar}{\tau}\left(\frac{\hbar}{\tau}+\frac{\hbar}{\tau_i}\right)}} \tag{2},$$

$$C = \frac{1}{1+\frac{2J(k)^2\left(1+\frac{\tau}{\tau_i}\right)}{2J(k)^2+\left(\frac{\hbar}{\tau}+\frac{\hbar}{\tau_o}\right)\left(\frac{\hbar}{\tau}+\frac{\hbar}{\tau_i}\right)}} \tag{3}.$$

Here, $\tau$ is the population decay time; $\tau_i$ and $\tau_o$ are in-plane and out-of-plane pseudospin scattering times respectively [$J(k)$ is the exchange interaction strength as defined in Eqn.(1)].

We first consider temperature dependence below 50K with isotropic scattering assumption $\tau_i = \tau_o = \tau_*$ in line with the original MSS model. Quantitative understanding of our data in the frameworks of Eqns.1&2 requires knowledge of $J(k)$ and $\tau$. Although their accurate values are difficult to assess directly, our experimental results place strong bounds on their possible ranges. At base temperature, the small *P* indicates that the exchange interaction is very efficient in depolarizing optically generated valley excitons, suggesting a clean/weak-scattering regime [30,40]. This limits the duration over which the excitons experience the exchange interaction to be comparable to or even longer than $\Omega^{-1}$. On the other hand, this time is not expected to be much longer than $\Omega^{-1}$, since a moderate temperature increase can raise $P$ quickly and drive the system into the strong scattering regime. At a minimum, the bright exciton must experience the exchange interaction in the light cone during its radiative lifetime. The exchange interaction in the light cone averages to about 0.5meV (Supplementary [27] S3). It turns out the radiative lifetime of ~2-4 ps [41–44] is already comparable to $\Omega^{-1}$. This suggests that excitons responsible for the radiative emission of PL must lose their initial 17 meV kinetic energy and relax towards the light cone rapidly. We comment that this conjecture is consistent with time-resolved PL measurements where the exciton emission occurs almost immediately after optical pumping [41–44], and is substantiated by the insensitivity of $P$ to changes of excitation photon energy in our PLE studies (Supplementary [27] Fig.S5). Within these constraints, we tested several different pairs of $J(k)$ and $\tau$ to fit our data (Supplementary [27] S4 and Fig.S2). In Fig.3c, we show least square fit using $J(k) = 1$ meV and $\tau = 5$ ps. With only one variable $\tau_*$ changing from ~0.05–1ps, the simulation captures our T < 50K data (symbols) well: both $P$ and $C$ increase at shorter $\tau_*$, reflecting scattering enhanced valley performance; also, $C$ is larger than $P$, consistent with the in-plane nature of the effective exchange field [25].

The same model can be used to fit the gate dependent $P$ and $C$. Here, $\tau_i$ is not equal to $\tau_o$ due to anisotropic pseudospin scattering as explained above. In particular, we anticipate that charge doping leads to much higher scattering rate for in-plane pseudospin, i.e. $\tau_i$ decreases much faster than $\tau_o$ due to the intervalley nature of exciton polaron dressing. In Fig.3d, we have plotted the extreme case where we keep $\tau_o$ a constant ~0.6 ps ($\frac{\hbar}{\tau_o} = 1.1$ meV) and let $\tau_i$ vary. As $\tau_i$ decreases, $P$ increases rapidly similar to the behavior of $P$ in Fig.3c, while $C$ maintains a low value and decreases slightly, consistent with experimental data (open symbols).

Simulations in Figs. 3c&3d capture the most salient features of our key experimental findings. Namely, for isotropic scattering both $P$ and $C$ are improved, while for anisotropic scattering only $P$ but not $C$ is enhanced. Both types of behavior are well described within a unified framework of our generalized MSS model that involves a minimal set of fitting parameters in Eqns.1&2. We further comment that the behavior of our device represents a transition from weak scattering to strong scattering regime, and for this reason, marked changes in valley polarization are observed. As an independent confirmation, we also measured T and Vg dependent $P$&$C$ in 1L-$WSe_2$ (Supplementary [27] S8). Similar trends were observed, and this verifies that scattering induced valley polarization enhancement is not limited to $MoS_2$ only.

With the above understanding of scattering enhanced valley polarization, we can simulate our experimental T and $V_g$ dependence of P&C. The high temperature dropping of valley polarization $P$ is not due to exchange interaction, and has been modeled before with other thermally activated intervalley scattering processes assisted by modes such as K-point phonons [5]. To obtain analytic expressions that can fit *P&C* over the whole temperature range, we revised our model in Supplementary [27] S2&S4 incorporating an additional parameter $\tau_b$ accounting for valley depolarization processes beyond exchange interaction. With this more comprehensive model, we arrive at Eqns.S12&S13, with which we successfully reproduced the T dependence of P & C over the whole temperature range (smooth curves in Fig.3a). For gate dependence, we fit our data employing in Eqns.(1&2) an empirical gate dependence $\frac{\hbar}{\tau_i} \approx a_0 + a_2(V_g - V_{g0})^2$, where $a_0$ = 0.8 meV, and $a_2$ = 6.4 meV/V$^2$ ($\frac{\hbar}{\tau_i}$ is treated as constant $a_0$ for $V_g < V_{g0}$ = 0.48 V) while keeping $\tau_o$ a constant. The simulation (Fig. 3b smooth curves) provides good match to our data (symbols).

In conclusion, our work presents an unusual phenomenon of scattering induced TMD valley polarization enhancement. The observed effects leave little doubt regarding the role of exchange interactions in 1L-TMD exciton valleytronics. The generalized MSS model developed here captures essential features of both valley polarization and coherence with respect to controlled scattering introduced via thermal activation and charge doping, in which the transition from weak to strong scattering is observed. These results shed key light to valley phenomena in two-dimensional TMDs, such as why cleaner TMD devices do not necessarily show better valley polarization properties.

**Acknowledgments**

This work is supported by National Science Foundation (NSF Grant number: DMR-2004474). K.W. and T.T. acknowledge support from the Elemental Strategy Initiative conducted by the Japan Ministry of Education, Culture, Sports, Science and Technology (MEXT, Grant Number: JPMXP0112101001), the Japan Society for the Promotion of Science Grants-in-Aid for Scientific Research (JSPSKAKENHI Grant Numbers JP20H00354) and the Japan Science and Technology Agency Core Research for Evolutional Science and Technology (CREST, Grant: JPMJCR15F3).

**References**

[1] A. Rycerz, J. Tworzydło, and C. W. J. Beenakker, Nat. Phys. **3**, 172 (2007).

[2] D. Xiao, W. Yao, and Q. Niu, Phys. Rev. Lett. **99**, 236809 (2007).

[3] D. Xiao, G.-B. Liu, W. Feng, X. Xu, and W. Yao, Phys. Rev. Lett. **108**, 196802 (2012).

[4] T. Cao, G. Wang, W. Han, H. Ye, C. Zhu, J. Shi, Q. Niu, P. Tan, E. Wang, B. Liu, and J. Feng, Nat. Commun. **3**, 887 (2012).

[5] H. Zeng, J. Dai, W. Yao, D. Xiao, and X. Cui, Nat. Nanotechnol. **7**, 490 (2012).

[6] K. F. Mak, K. He, J. Shan, and T. F. Heinz, Nat. Nanotechnol. **7**, 494 (2012).

[7] G. Kioseoglou, A. T. Hanbicki, M. Currie, A. L. Friedman, and B. T. Jonker, Sci. Rep. **6**, 25041 (2016).

[8] G. Kioseoglou, A. T. Hanbicki, M. Currie, A. L. Friedman, D. Gunlycke, and B. T. Jonker, Appl. Phys. Lett. **101**, 221907 (2012).

[9] T. Yu and M. W. Wu, Phys. Rev. B **89**, 205303 (2014).

[10] M. Z. Maialle, E. A. de Andrada e Silva, and L. J. Sham, Phys. Rev. B **47**, 15776 (1993).

[11] H. Yu, G.-B. Liu, P. Gong, X. Xu, and W. Yao, Nat. Commun. **5**, 3876 (2014).

[12] H. Tornatzky, A.-M. M. Kaulitz, and J. Maultzsch, Phys. Rev. Lett. **121**, 167401 (2018).

[13] T. Yan, X. Qiao, P. Tan, and X. Zhang, Sci. Rep. **5**, 1 (2015).

[14] G. Sallen, L. Bouet, X. Marie, G. Wang, C. R. Zhu, W. P. Han, Y. Lu, P. H. Tan, T. Amand, B. L. Liu, and B. Urbaszek, Phys. Rev. B **86**, 81301 (2012).

[15] M. Yang, C. Robert, Z. Lu, D. Van Tuan, D. Smirnov, X. Marie, and H. Dery, Phys. Rev. B **101**, 115307 (2020).

[16] M. Selig, F. Katsch, S. Brem, G. F. Mkrtchian, E. Malic, and A. Knorr, Phys. Rev. Res. **2**, 23322 (2020).

[17] S.-Y. Chen, T. Goldstein, T. Taniguchi, K. Watanabe, and J. Yan, Nat. Commun. **9**, 3717 (2018).

[18] S.-Y. Chen, Z. Lu, T. Goldstein, J. Tong, A. Chaves, J. Kunstmann, L. S. R. Cavalcante, T. Woźniak, G. Seifert, D. R. Reichman, T. Taniguchi, K. Watanabe, D. Smirnov, and J. Yan, Nano Lett. **19**, 2464 (2019).

[19] Y.-C. Wu, S. Samudrala, A. McClung, T. Taniguchi, K. Watanabe, A. Arbabi, and J. Yan, ACS Nano (2020).

[20] F. Cadiz, E. Courtade, C. Robert, G. Wang, Y. Shen, H. Cai, T. Taniguchi, K. Watanabe, H. Carrere, D. Lagarde, M. Manca, T. Amand, P. Renucci, S. Tongay, X. Marie, and B. Urbaszek, Phys. Rev. X **7**, 021026 (2017).

[21] T. Jakubczyk, G. Nayak, L. Scarpelli, W.-L. Liu, S. Dubey, N. Bendiab, L. Marty, T. Taniguchi, K. Watanabe, F. Masia, G. Nogues, J. Coraux, W. Langbein, J. Renard, V. Bouchiat, and J. Kasprzak, ACS Nano **13**, 3500 (2019).

[22] C. Robert, B. Han, P. Kapuscinski, A. Delhomme, C. Faugeras, T. Amand, M. R. Molas, M.

Bartos, K. Watanabe, T. Taniguchi, B. Urbaszek, M. Potemski, and X. Marie, Nat. Commun. **11**, 4037 (2020).

[23] K. Hao, G. Moody, F. Wu, C. K. Dass, L. Xu, C.-H. Chen, L. Sun, M.-Y. Li, L.-J. Li, A. H. MacDonald, and X. Li, Nat. Phys. **12**, 677 (2016).

[24] G. Moody, K. Tran, X. Lu, T. Autry, J. M. Fraser, R. P. Mirin, L. Yang, X. Li, and K. L. Silverman, Phys. Rev. Lett. **121**, 57403 (2018).

[25] S.-Y. Chen, T. Goldstein, J. Tong, T. Taniguchi, K. Watanabe, and J. Yan, Phys. Rev. Lett. **120**, 46402 (2018).

[26] A. M. Jones, H. Yu, N. J. Ghimire, S. Wu, G. Aivazian, J. S. Ross, B. Zhao, J. Yan, D. G. Mandrus, D. Xiao, W. Yao, and X. Xu, Nat. Nanotechnol. **8**, 634 (2013).

[27] (n.d.).

[28] H. Tornatzky, R. Gillen, H. Uchiyama, and J. Maultzsch, Phys. Rev. B **99**, 144309 (2019).

[29] M. L. Lin, Q. H. Tan, J. Bin Wu, X. S. Chen, J. H. Wang, Y. H. Pan, X. Zhang, X. Cong, J. Zhang, W. Ji, P. A. Hu, K. H. Liu, and P. H. Tan, ACS Nano **12**, 8770 (2018).

[30] N. Bloembergen, E. M. Purcell, and R. V Pound, Phys. Rev. **73**, 679 (1948).

[31] M. I. D'yakonov and V. I. Perel', Sov. Phys. Solid State **13**, 3023 (1972).

[32] K. Shinokita, X. Wang, Y. Miyauchi, K. Watanabe, T. Taniguchi, and K. Matsuda, Adv. Funct. Mater. **29**, 1900260 (2019).

[33] D. K. Efimkin and A. H. MacDonald, Phys. Rev. B **95**, 035417 (2017).

[34] D. K. Efimkin and A. H. MacDonald, Phys. Rev. B **97**, 235432 (2018).

[35] M. Sidler, P. Back, O. Cotlet, A. Srivastava, T. Fink, M. Kroner, E. Demler, and A. Imamoglu, Nat. Phys. **13**, 255 (2017).

[36] J. G. Roch, G. Froehlicher, N. Leisgang, P. Makk, K. Watanabe, T. Taniguchi, and R. J. Warburton, Nat. Nanotechnol. **14**, 432 (2019).

[37] T. Goldstein, Y.-C. Wu, S.-Y. Chen, T. Taniguchi, K. Watanabe, K. Varga, and J. Yan, J. Chem. Phys. **153**, 71101 (2020).

[38] R. A. Sergeev and R. A. Suris, Phys. Status Solidi **227**, 387 (2001).

[39] J. Yan and K. Varga, Phys. Rev. B **101**, 235435 (2020).

[40] H. Yu, X. Cui, X. Xu, and W. Yao, Natl. Sci. Rev. **2**, 57 (2015).

[41] C. Robert, D. Lagarde, F. Cadiz, G. Wang, B. Lassagne, T. Amand, A. Balocchi, P. Renucci, S. Tongay, B. Urbaszek, and X. Marie, Phys. Rev. B **93**, 205423 (2016).

[42] D. Lagarde, L. Bouet, X. Marie, C. R. Zhu, B. L. Liu, T. Amand, P. H. Tan, and B. Urbaszek, Phys. Rev. Lett. **112**, 047401 (2014).

[43] G. Wang, L. Bouet, D. Lagarde, M. Vidal, A. Balocchi, T. Amand, X. Marie, and B. Urbaszek, Phys. Rev. B **90**, 75413 (2014).

[44] S.-Y. Chen, M. Pieczarka, M. Wurdack, E. Estrecho, T. Taniguchi, K. Watanabe, J. Yan, E. A.

Ostrovskaya, and M. S. Fuhrer, ArXiv:2009.09602 (2020).

[45] G. Wang, E. Palleau, T. Amand, S. Tongay, X. Marie, and B. Urbaszek, Appl. Phys. Lett. **106**, 112101 (2015).

[46] D. MacNeill, C. Heikes, K. F. Mak, Z. Anderson, A. Kormányos, V. Zólyomi, J. Park, and D. C. Ralph, Phys. Rev. Lett. **114**, 037401 (2015).

[47] T. Jakubczyk, K. Nogajewski, M. R. Molas, M. Bartos, W. Langbein, M. Potemski, and J. Kasprzak, 2D Mater. **5**, 031007 (2018).

[48] Y. Miyauchi, S. Konabe, F. Wang, W. Zhang, A. Hwang, Y. Hasegawa, L. Zhou, S. Mouri, M. Toh, G. Eda, and K. Matsuda, Nat. Commun. **9**, 2598 (2018).

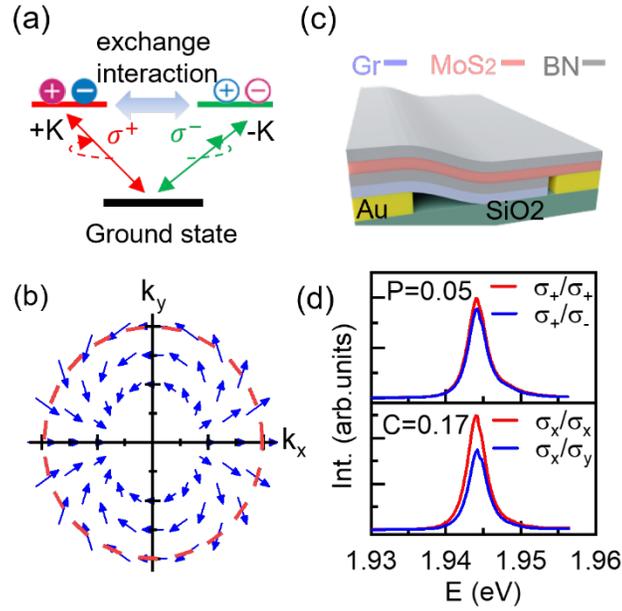

FIG. 1. (a) The optical selection rule and exchange interaction for excitons in +K and -K valleys of monolayer $MoS_2$. (b) The strength and direction of the intervalley exchange pseudomagnetic field in $k$ space. (c) Schematic drawing of the gated hBN-encapsulated $MoS_2$ device. (d) Polarization-resolved PL of 1L-$MoS_2$ at $V_g$=0.

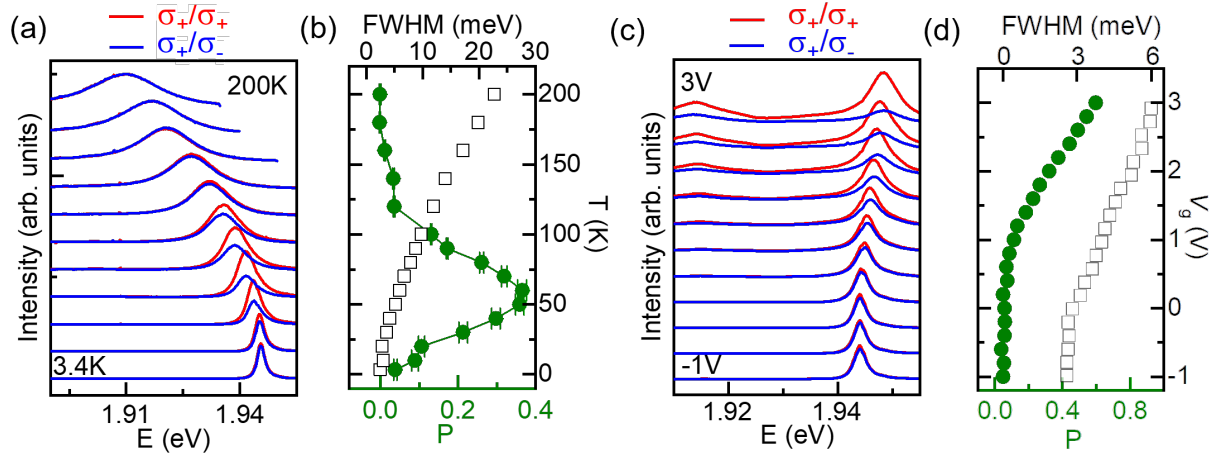

FIG. 2. (a) Temperature dependence of circular-polarization resolved PL at $V_g = -1$V. (b) Valley polarization (solid) and PL linewidth (hollow) extracted from data in (a). (c) Gate voltage dependence of circular-polarization resolved PL at 3.4K. (d) Valley polarization (solid) and PL linewidth (hollow) extracted from data in (c).

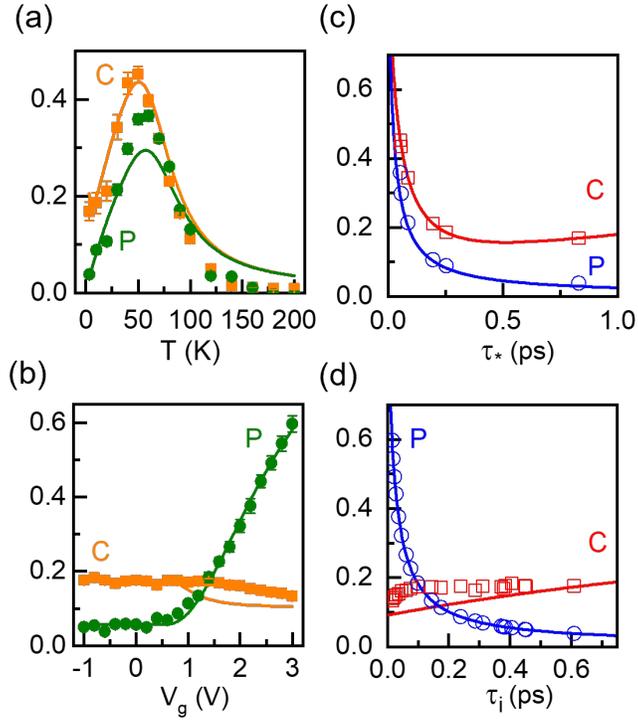

FIG. 3. (a) Temperature and (b) gate voltage dependence of valley coherence (squares) *vs.* polarization (circles). The smooth curves are theoretical fittings. (c) Fitting of pseudospin scattering time for data in (a) below 50K. (d) Fitting of pseudospin scattering time for $V_g$ dependent data in (b).